\documentclass[prb, twocolumn, nofootinbib]{revtex4-2}
\usepackage{bm}
\usepackage{graphicx}
\usepackage[x11names]{xcolor} 
\usepackage{amsmath}
\usepackage{bbold}
\usepackage{mathrsfs}
\usepackage{color}
\usepackage[unicode=true,colorlinks=true]{hyperref}
\usepackage[capitalize]{cleveref}

\begin{document}

\title{Application of perturbation theory to finding vibrational frequencies of a spheroid}
\author{M.O.~Nestoklon} 
\affiliation{Experimentelle Physik 2, Technische Universit\"at Dortmund, 44221 Dortmund, Germany}
\author{L.~Saviot}
\affiliation{Universit\'e Bourgogne Europe, CNRS, Laboratoire Interdisciplinaire Carnot de Bourgogne
ICB UMR 6303, 21000 Dijon, France}
\author{S.V.~Goupalov} 
\email{serguei.goupalov@jsums.edu}
\affiliation{Department of Physics, Jackson State University, Jackson MS 39217, USA}

\begin{abstract}
We apply perturbation theory of boundary conditions, originally developed by A.B.~Migdal and independently by S.A.~Moszkowski for deformed atomic nuclei, to finding eigenfrequencies of Raman-active
spheroidal modes of a spheroid from these of a sphere and compare the outcomes with the results of numerical calculations for CdSe and silver nanoparticles. The modes are characterized by the total angular momentum $j=2$ and are five-fold degenerate for a sphere but split into three distinct modes, characterized by the absolute value of the total angular momentum projection onto the spheroidal axis, in case of a spheroid. The perturbation method works well in case of the rigid boundary
conditions, with the displacement field set to zero at the boundary, and accurately predicts the splittings when the spheroidal shape is close to a sphere, but fails in case of the stress-free boundary conditions.
\end{abstract}

\maketitle

\section{Introduction} 
Acoustic vibrations of semiconductor nanocrystals are responsible for various relaxation and dephasing processes in these
materials and determine broadening of spectral lines in their absorption and photoluminescence spectra~\cite{suris,han}. Vibrational modes of semiconductor and metal
nanoparticles can be probed directly in
low-frequency Raman spectroscopy~\cite{duval86,saviot98,ekimov,saviot_review,harkort} and time-domain transient absorption experiments~\cite{krauss,hartland11,hartland25}.

Comparison with atomistic calculations has revealed that the frequencies of fundamental Raman-active vibrational modes can be found within the approximation of elastic continuum for nanocrystals containing as few as 20 atoms~\cite{combe}. In case of elastically isotropic nanocrystals of spherical shape, the frequencies of free vibrations can be found analytically. The solution, involving only trigonometric functions, was obtained in the last quarter of XIX century~\cite{jaerisch,lamb1882,chree}.

For more complex nanocrystal shapes, numerical procedures of finding the frequencies of free-standing vibrations have been developed~\cite{visscher,saviot21,pelton}.
However, it is quite desirable to extend the analytical treatment beyond the case of a sphere. In particular, this would provide a natural way to
check the numerical results. 

A slightly imperfect sphere can often be described as an ellipsoid. A uniform uniaxial deformation of a sphere yields a spheroid. When this deformation 
is small, a parameter of non-sphericity can be introduced, and corrections to the vibrational frequencies, linear in this parameter, can be sought.
Then a more complex shape of a three-axial ellipsoid can be treated as a superposition of uniaxial deformations along perpendicular axes.
Vibrational modes of spheroidal (or almost spheroidal) nanoparticles are also of interest in their own right, as they have been probed in experiments on low-frequency Raman scattering~\cite{mariotto,margueritat}.

Spheroidal shapes can be treated perturbatively using the method known as the perturbation theory of the boundary conditions, originally developed by A.B.~Migdal~\cite{migdal,ll} and independently by S.A.~Moszkowski~\cite{moszkowski} to find energy levels of a deformed atomic nucleus.
In this method, a spheroidal shape results from a uniform uniaxial deformation of a sphere preserving its volume. Performing a coordinate transformation, one can regain the spherically symmetric boundary condition at the expense of an anisotropic addition to the Hamiltonian, treated as a perturbation. 

The method seems to be bulletproof in case of the rigid boundary conditions, when the vector field of the displacement is set to be zero at the boundary of the spheroid. However, in case of the free boundary
conditions, when traction is set to be zero at the boundary, it is not evident that the stress-free sphere in the transformed coordinates would adequately represent a stress-free spheroid in the initial
coordinate system. As the formulation of the method is essentially independent of the particular form of boundary conditions, here we will apply it to both cases and compare the outcomes to the results of numerical solutions. We note that the rigid boundary conditions are often used when discussing the influence of a matrix surrounding nanocrystals, {\it e.g.} in the case of nanocrystals embedded in glass~\cite{saviot96}. In practice, the actual frequencies for a nanoparticle embedded in a solid matrix are often intermediate between these yielded by the free and rigid boundary conditions.

Consider a spheroidal nanocrystal with the surface 
\begin{equation}
\frac{x^2+y^2}{b^2}+\frac{z^2}{c^2}=1 \,.
\end{equation}
Substitution $x \rightarrow bx/R$, $y \rightarrow by/R$, $z \rightarrow cz/R$ transforms it into a sphere $x^2+y^2+z^2=R^2$. Introducing the ellipsoidality parameter
\begin{equation}
\mu_z=2 \frac{c-b}{c+b} \,,
\end{equation}
(in Migdal's book and Moszkowski's article the notations $\beta$ and $3 \, d/2$, respectively, are used for $\mu_z$)
and assuming $|\mu_z| \ll 1$, from the condition $c \,b^2=R^3$ one obtains
\[
b \approx R \left(1 -\frac{\mu_z}{3} \right) \,, \,\,\,\,\,\, c \approx R \left(1 +\frac{2\mu_z}{3} \right) \,.
\] 
Application of the transformation $x \rightarrow x \, (1 -\mu_z/3)$, $y \rightarrow y \, (1 -\mu_z/3)$, $z \rightarrow z \, (1 +2 \mu_z/3)$
to the system's Hamiltonian, or rather its kinetic energy part, yields the linear in $\mu_z$ perturbation.


Migdal's theory is linear in $\mu_z$ and valid for $|\mu_z| \ll 1$, although higher-order corrections can also be considered~\cite{moszkowski}. 

This perturbation theory was applied to finding valence-band states in spheroidal nanocrystals of semiconductors with strong spin-orbit coupling. In bulk semiconductors of III-V and II-VI compounds the states near the 
valence-band maximum are described by the effective Luttinger Hamiltonian which in the isotropic limit takes the form~\cite{luttinger}
\begin{equation}
\hat{H}({\bf k})=\frac{\hbar^2}{2 m_0} \left[- \left(\gamma_1 + \frac{5}{2} \, \gamma \right)  k^2+2 \, \gamma \left({\bf kJ} \right)^2 \right] \,,
\label{lut}
\end{equation}
where $\hbar$ is Planck's constant, $m_0$ is the free-electron mass, $\gamma=(2 \, \gamma_2+3 \, \gamma_3)/5$; $\gamma_1$, $\gamma_2$, and $\gamma_3$ are the Luttinger parameters, ${\bf k}$ is the quasi-wave vector, and $J_{\alpha}$ ($\alpha=x,y,z$) are $4 \times 4$ matrices corresponding to the spin $3/2$. When one wants to also account for the spin-orbitally split-off valence band, a modification of this Hamiltonian is used~\cite{dp_jetp,apy,dp}
\begin{equation}
\hat{H}({\bf k})=-\frac{\hbar^2 k^2}{2 m_0} (\gamma_1+ 4 \gamma) +\frac{3 \hbar^2 \gamma}{m_0} ({\bf k} {\bf S} )^2 + \frac{\Delta}{3} ({\bm \sigma} {\bf S} ) -\frac{\Delta}{3} \,, 
\label{lutgen}
\end{equation}
where  $S_{\alpha}$ ($\alpha=x,y,z$) are the matrices of angular momentum $S=1$, $\Delta$ is the spin-orbit splitting, and $\sigma_{\alpha}$ are the Pauli matrices.
The methods of constructing confined electronic states from the bands, described by the Hamiltonians~(\ref{lut}) and~(\ref{lutgen}), in a spherical nanocrystal were developed in Refs.~\cite{sercel} and~\cite{richard},
respectively, while the perturbation theory of the boundary conditions, accounting for spheroidal shapes, was applied to these models in Refs.~\cite{efroro} and~\cite{shape3b}, respectively.

Equations of motion for an isotropic elastic medium can be recast~\cite{merkulov,ufn} as an eigenvalue problem with an operator very similar to Eqs.~(\ref{lut}) and~(\ref{lutgen}):
\begin{equation}
\hat{\Lambda}(-i \, \bm{\nabla}) \bm{u}=\omega^2 \, \bm{u} \,,
\end{equation}
\begin{equation}
\hat{\Lambda}({\bf k})=-(c_l^2-c_t^2) ({\bf k} {\bf S})^2+ c_l^2 \, k^2   \,,
\label{isotrop}
\end{equation}
where $\bm{u}$ is the displacement field and $c_l$ and $c_t$ are the longitudinal and transverse sound velocities. The solutions of this equation describing vibrational modes of a free-standing or restrained spherical nanoparticle can be constructed~\cite{ufn} in the same way
as the valence band states in a spherical semiconductor nanocrystal~\cite{sercel,richard}. Likewise, application of the perturbation theory of the boundary conditions in order to account for the spheroidal shape of a nanocrystal, should 
also be quite similar. We will consider its application to the Raman-active~\cite{duval92} spheroidal mode $_nS_2^m$ of a spherical nanoparticle.
Other attempts to adapt the perturbation theory of boundary conditions to the problems of the theory of elasticity we are aware of~\cite{tamura} did not take into account the spin structure of Eq.~(\ref{isotrop}).

\section{Spheroidal vibrations of a spherical nanoparticle}

The displacement field for the spheroidal phonon mode characterized by the total angular momentum $j=2$ and its projection $m$ is given by~\cite{ufn,nestoklon}
\begin{widetext}
\begin{multline}
\label{solution}
\langle {\bf r}|_nS_2^m \rangle=A \, R^{-3/2} \, 
\Bigg\{
\left[ \sqrt{\frac{3}{5}} \, j_3(q_nr) \, {\bf Y}^3_{2m} \left(\frac{{\bm r}}{r} \right) +
\sqrt{\frac{2}{5}} \, j_1(q_nr) \, {\bf Y}^1_{2m} \left(\frac{{\bm r}}{r} \right) \right]
\\
+ \frac{B}{A} \, \left[ \sqrt{\frac{2}{5}} \, j_3(Q_nr) \, {\bf Y}^3_{2m} \left(\frac{{\bm r}}{r} \right) -
\sqrt{\frac{3}{5}} \, j_1(Q_nr) \, {\bf Y}^1_{2m} \left(\frac{{\bm r}}{r} \right) \right] 
\Bigg\}
\,,
\end{multline}
where
\begin{align*}
\frac{B}{A}&=\frac{5 \, c_l^2 \,q_nR \, j_2(q_nR) - 4 \, c_t^2 \, j_1(q_nR)-24 \, c_t^2 \, j_3(q_nR)}{2 \, c_t^2 \, \sqrt{6} \, \left(4 \, j_3(Q_nR)-j_1(Q_nR) \right)}
\end{align*}
for the free boundary conditions and
\begin{align*}
\frac{B}{A}&=\frac{2 \, j_1(q_nR) - 3 j_3(q_nR)}{\sqrt{6} \, \left(j_1(Q_nR)+j_3(Q_nR) \right)}
\end{align*}
\end{widetext}
for the rigid boundary conditions,
$j_l(x)$ are the spherical Bessel functions, ${\bf Y}^l_{jm}({\bf r}/r)$ are the vector spherical harmonics~\cite{vmk}, the wave numbers $q_n$ and $Q_n$ are related to the five-fold 
degenerate mode frequency $\omega_n$ through $q_n=\omega_n/c_l$, $Q_n=\omega_n/c_t$, and the frequency is determined as the $(n+1)$-th root of the transcendental equation~\cite{saviot96,ufn}
\begin{equation}
j_2(q_nR) \, j_2(Q_nR) \, (Q_nR)^2 \, \left[5-\frac{(Q_nR)^2}{2} \right]
\label{spheroid}
\end{equation}
\[
+8 \, j_3(q_nR) \, j_3(Q_nR) \,(q_nR) \, (Q_nR) 
\]
\[
+j_2(q_nR) \, j_3(Q_nR) \, Q_nR \, \left[ (Q_nR)^2-16 \right]
\]
\[
+2 \, j_3(q_nR) \, j_2(Q_nR) \,q_nR \, \left[ (Q_nR)^2-12 \right] =0 
\]
for the free boundary conditions and
\begin{equation}
3 \, j_3(q_nR) \, j_1(Q_nR) +2 \, j_3(Q_nR) \, j_1(q_nR)=0
\label{spheroid_rigid}
\end{equation}
for the rigid boundary conditions.

The details of the procedure of second quantization for this mode can be found in Ref.~\cite{nestoklon}. For the purpose of the present study we will normalize the displacement field by the condition
\[
\langle _{n'}S_2^{m'}|_nS_2^m \rangle=\delta_{n,n'} \, \delta_{m,m'}
\]
which yields an equation for $A$:
\begin{widetext}
\begin{multline}
A^2 \, \Bigg\{ \int\limits_0^1 dx \, x^2 \left[ 3 j_3^2(q_n R x) +2 \, j_1^2(q_n R x) \right] 
+\frac{B^2}{A^2} \, \int\limits_0^1 dx \, x^2 \left[ 2 j_3^2(Q_n R x) +3 \, j_1^2(Q_n R x) \right]
\\
+2 \sqrt{6} \frac{B}{A} \int\limits_0^1 dx \, x^2 \left[ j_3(q_n R x) \, j_3(Q_n R x) 
- j_1(q_n R x) \, j_1(Q_n R x) \right]
\Bigg\}=5 \,.
\end{multline}
\end{widetext}

\section{Shape anisotropy}

We shall apply the transformation  $k_{x,y} \rightarrow k_{x,y} \, (1+\mu_z/3)$, $k_z \rightarrow k_z \, (1-2 \mu_z/3)$ to the operator~(\ref{isotrop}). This yields the following linear in $\mu_z$ addition to
this operator:
\begin{widetext}
\begin{equation}
\label{anisotrop}
\Delta \hat{\Lambda}(-i \, \bm{\nabla})=\mu_z \, \left\{
\frac{2}{3} (c_l^2-c_t^2) \, ({\bf S} \bm{\nabla} )^2 - (c_l^2-c_t^2) \, \left[ ({\bf S} \bm{\nabla}) S_z \nabla_z+  S_z \nabla_z ({\bf S} \bm{\nabla}) \right] -
\frac{2}{3} \, c_l^2 \, \bm{\nabla}^2 + 2 c_l^2 \nabla_z^2 \right\} \,.
\end{equation}
\end{widetext}
Calculation of the matrix elements of this operator on the solutions~(\ref{solution}) with various values of $m$ should lead to the matrix in the form $C_1+C_2 \, (J_z^2-2)$, where 
$J_z$ is the matrix of the $z$-projection of the angular momentum $j=2$ and $C_1$ and $C_2$ are real coefficients. From symmetry considerations $C_1=0$. But the term proportional to
$(J_z^2-2)$ can only originate from the terms in Eq.~(\ref{anisotrop}) explicitly containing $\nabla_z$. Therefore, we will calculate matrix elements of the terms in Eq.~(\ref{anisotrop}) explicitly 
containing $\nabla_z$ and disregard the 
terms proportional to the unit matrix in the result. 

The relevant matrix elements on $\langle {\bf r}|_nS_2^m \rangle$ are diagonal and, using the machinery outlined in~\cite{vmk},  can be represented in the form
\begin{widetext}
\begin{equation}
\left\langle \left. \left. _nS_2^{m'} \right| (c_t^2-c_l^2) \, \left[ ({\bf S} \bm{\nabla}) S_z \nabla_z+  S_z \nabla_z ({\bf S} \bm{\nabla}) \right] 
+ 2 \, c_l^2 \, \nabla_z^2 \right| \protect{_nS_2^m} \right\rangle=\delta_{m,m'} \, \sum\limits_{l=1,2,3} d_{l,n} \, \left[
\left(
\begin{matrix}
2&1&l\cr
m & 0&-m
\end{matrix}
\right)
\right]^2
\,,
\end{equation}
\end{widetext}
where $\left(
\begin{matrix}
2&1&l\cr
m & 0&-m
\end{matrix}
\right)
$ is the Wigner 3jm-symbol~\cite{vmk} and $d_{l,n}$ is a coefficient.

The resulting $C_2$ coefficient is given by
\begin{widetext}
\begin{equation}
\label{C2}
C_2=\frac{\mu_z \, A^2 \, R^{-3}}{5 \cdot 7} \, \int\limits_0^R dr \, r^2 \, \left\{
(c_t^2-c_l^2) \, \left[ q_n^2 \, j_3^2 (q_nr) +\frac{7}{3} \, Q_n^2 \, \frac{B^2}{A^2} \, j_3^2 (Q_nr)
+\frac{7 \sqrt{6}}{6} \, Q_n^2 \, \frac{B}{A} \, j_3(q_nr) \, j_3(Q_nr)
\right. \right.
\end{equation}
\[
\left. \left.
+\frac{\sqrt{6}}{3} \, q_n^2 \, \frac{B}{A} \, j_3(q_nr) \, j_3(Q_nr) -q_n^2 \, j_1^2(q_nr)
+ Q_n^2 \, \frac{B^2}{A^2} \, j_1^2 (Q_nr)
+\frac{\sqrt{6}}{2} \, q_n^2 \, \frac{B}{A} \, j_1(q_nr) \, j_1(Q_nr)
-\frac{\sqrt{6}}{3} \, Q_n^2 \, \frac{B}{A} \, j_1(q_nr) \, j_1(Q_nr) \right]
\right.
\]
\[
\left.
+2 c_l^2 \left[
2 \, q_n^2 \, j_3^2 (q_nr) +\frac{2}{3} \, Q_n^2 \, \frac{B^2}{A^2} \, j_3^2 (Q_nr)
+\frac{2}{3} \, \sqrt{6} \, q_n^2 \, \frac{B}{A} \, j_3(q_nr) \, j_3(Q_nr)
+\frac{\sqrt{6}}{3} \, Q_n^2 \, \frac{B}{A} \, j_3(q_nr) \, j_3(Q_nr) +\frac{4}{3} \,q_n^2 \, j_1^2(q_nr)
\right. \right.
\]
\[
\left. \left.
+ Q_n^2 \, \frac{B^2}{A^2} \, j_1^2 (Q_nr)
-\frac{2 \sqrt{6}}{3} \, q_n^2 \, \frac{B}{A} \, j_1(q_nr) \, j_1(Q_nr) 
- \frac{\sqrt{6}}{3} \, Q_n^2 \, \frac{B}{A} \, j_1(q_nr) \, j_1(Q_nr) \right] \right\} \,.
\]
\end{widetext}
The matrix $C_2 \, (J_z^2-2)$ has degenerate eigenvalues of $2 C_2$ and $- C_2$ and a non-degenerate eigenvalue of $-2 C_2$ which represent linear in $\mu_z$ corrections to $\omega_n^2$.
The corresponding corrections to $\omega_n$ are $C_2/\omega_n$, $- C_2/(2 \, \omega_n)$, and $- C_2/\omega_n$. One can see from Eq.~(\ref{C2}) that $C_2 \propto R^{-2}$ while $\omega_n$ and,
therefore, $C_2/\omega_n \propto R^{-1}$.
  
\begin{figure*}[ht]
  \centering
    \includegraphics[width=\textwidth]{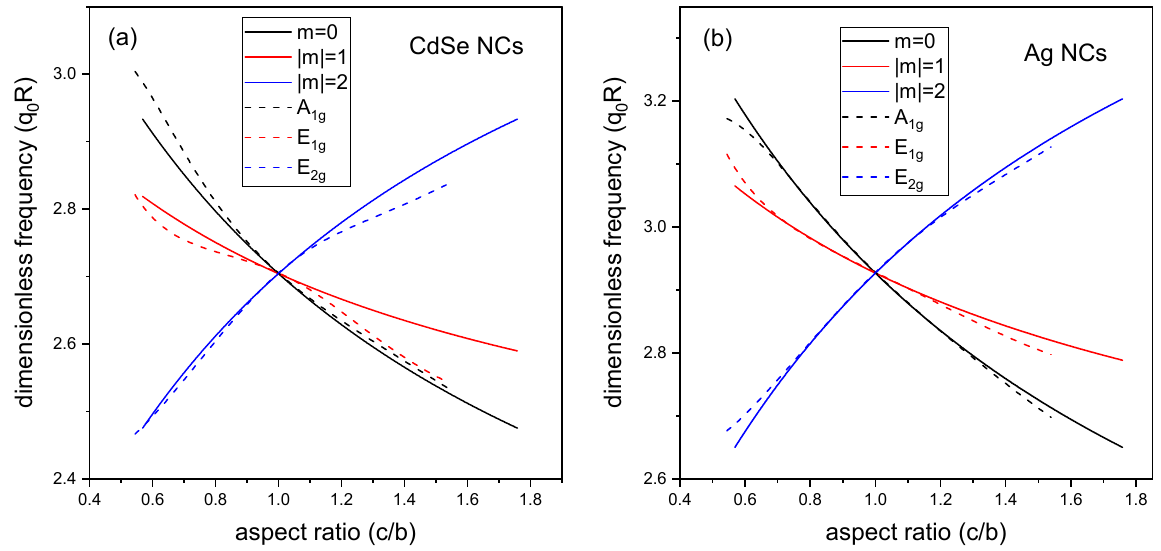}
\caption{(a): Dimensionless vibration frequencies ($q_0 R$) of the fundamental spheroidal mode with $j=2$ and $|m|=0,1,2$ as functions of the spheroid aspect ratio, $c/b$, for a restrained CdSe nanocrystal of mean radius $R=c^{1/3} \, b^{2/3}$.
The results of the perturbation theory (solid lines) are compared to numerical calculations (dashed lines). The rigid boundary conditions are used in both cases. Numerical calculations were performed for $R=3$~nm. 
(b): Same as (a) but for a silver nanocrystal. }
\label{fig1}
\end{figure*}

\begin{figure*}[ht]
  \centering
    \includegraphics[width=\textwidth]{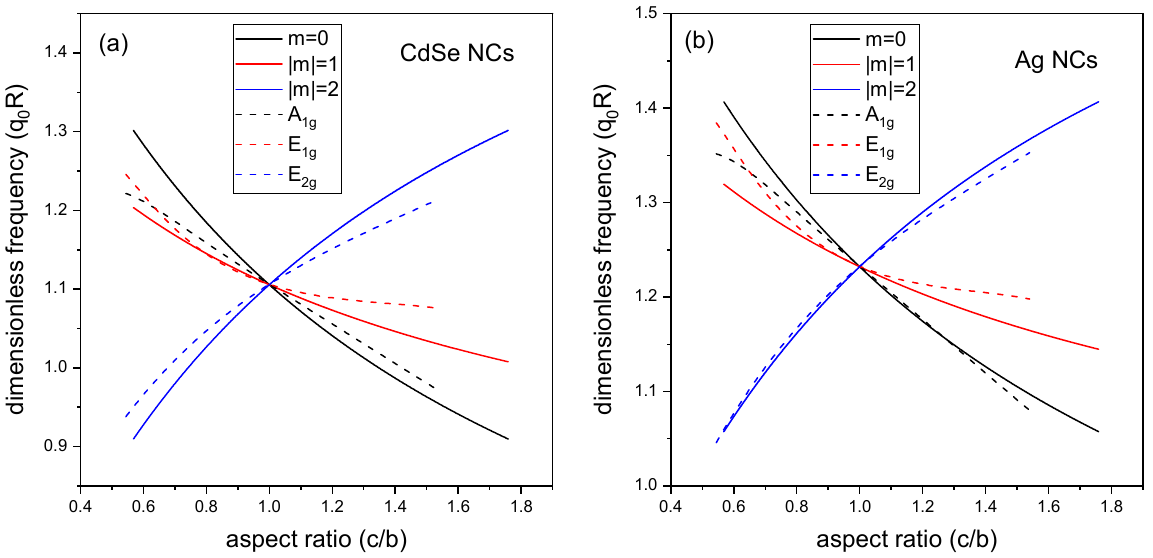}
\caption{Same as Fig.~\ref{fig1} but for the free boundary conditions.}
\label{fig2}
\end{figure*}

\section{Numerical calculations}

To test the validity of the perturbation theory approximation,
we calculated the vibrations with the free boundary condition using the Rayleigh-Ritz approach as described in Ref.~\onlinecite{visscher}.
It relies on the choice of products of powers of the Cartesian coordinates ($x^ly^mz^n$) as a basis for expansion of the displacement.
We used $l+m+n \le 20$ in this work. No additional restrictions on the basis functions are needed in the case of free boundary conditions, 
because they emerge as natural boundary conditions from the variational formulation~\cite{visscher,sagan,gould}.
This basis was modified for the rigid boundary conditions to satisfy the vanishing displacement at the surface of the spheroid by using $x^ly^mz^n \times (1-\frac{x^2}{b^2}-\frac{y^2}{b^2}-\frac{z^2}{c^2})$ functions.
These are linear combinations of $x^{l'}y^{m'}z^{n'}$ functions.
The integrals of products of the new basis functions over the volume of the spheroid can still be expressed analytically.
These are the main ingredients of the numerical calculations.
Therefore, the method is rather straightforward to implement and the benefits of the $x^ly^mz^n$ basis functions are maintained.

We took advantage of the $x$, $y$ and $z$ symmetry planes as in Ref.~\onlinecite{saviot21} to speed up the calculations and identify the irreducible representation associated with each vibration.
The point group for a spheroid made of a material with isotropic elasticity is D\textsubscript{$\infty$h}.
The calculations were performed for D\textsubscript{4h} because of the choice of Cartesian coordinates to express the basis functions.
The spheroidal phonon mode with total angular momentum $j=2$ of the sphere splits into the A\textsubscript{1g}, E\textsubscript{1g} and E\textsubscript{2g} irreducible representation in  D\textsubscript{$\infty$h} corresponding to $m=0$, $m=\pm1$ and $m=\pm2$ respectively.
These correspond to A\textsubscript{1g}, E\textsubscript{g} and B\textsubscript{1g}+B\textsubscript{2g} in D\textsubscript{4h} respectively.
The branches plotted in the following correspond to those of these irreducible representations which match the frequency of the spheroidal phonon mode with total angular momentum $j=2$ for the sphere.

For the rigid boundary conditions, the parity of the basis functions is still governed by $l$, $m$ and $n$ which means that the symmetry considerations used for the free boundary conditions remain valid.

The validity of the calculations was checked by comparing the obtained frequencies with those of the sphere as determined from Eq.~(\ref{spheroid}) or Eq.~(\ref{spheroid_rigid}).
The frequency for the sphere is reproduced with a precision of $10^{-8}$ with the Rayleigh-Ritz approach.
We also checked the calculations for spheroids with the finite element method for silver spheroids with $b/c=0.6$ and $b/c=1.5$
with the FreeFem++ software~\cite{freefem} using a mesh characteristic length 20 times smaller than the radius of the sphere.
The obtained frequencies for the modes of interest matched the Rayleigh-Ritz ones with a precision of $4\times10^{-4}$.

\section{Results and discussion}

In Fig.~\ref{fig1} we compare the results of the perturbation theory for CdSe and silver spheroids with mean radii $R=c^{1/3} \, b^{2/3}$ 
and the rigid boundary conditions with the results of the numerical calculations of vibrational eigenfrequencies. We use the following values of sound velocities:
$c_l(\mbox{CdSe})=3.7 \cdot 10^5$~cm/s, $c_t(\mbox{CdSe})=1.54 \cdot 10^5$~cm/s, $c_l(\mbox{Ag})=3.747 \cdot 10^5$~cm/s, $c_t(\mbox{Ag})=1.74 \cdot 10^5$~cm/s and restrict our consideration by the
fundamental spheroidal mode with $j=2$ ($_0S_2^m$). The numerical calculations were performed for $R=3$~nm.

One can see that in both cases of CdSe and Ag nanocrystals the two solutions match in the vicinity of the unit aspect ratio ($c/b=1$). Interestingly, this vicinity is substantially wider in case of 
silver nanocrystals as compared to CdSe nanocrystals. This is likely due to a closer proximity of the torsional mode with the total angular momentum $j=3$ ($_0T_3^m$) in the latter case.

The comparison for the free boundary conditions is presented in Fig.~\ref{fig2}.
While in case of silver nanocrystals the comparison of the numerical results and the outcomes of the perturbation theory is not too bad, Fig.~\ref{fig2},~a clearly shows that the perturbation theory
is inadequate for the free  boundary conditions. This points to the fact, that although the coordinate transform used in Migdal's theory preserves the volume, it does not preserve angles. As
a result, the stress-free sphere in the transformed coordinates does not correspond to a stress-free spheroid in the initial coordinate system, even as far as only linear terms in non-sphericity are concerned.

\section{Conclusions}

In this work we have applied the perturbation theory of the boundary conditions in order to find vibrational frequencies of a spheroid from these of a sphere. The perturbation procedure is based on a trick whereby the
coordinate change transforms a spheroidal surface to a spherical one of the same enclosed volume. Intuitively, we expect that the boundary condition at the surface, whether it corresponds
to a restrained or a stress-free one, formulated for the transformed (spherical) surface, would adequately represent the physical boundary conditions at the spheroidal one as well. However, mathematical 
requirements imposed on this transformation should be different in the two cases, and, for the latter one, the preservation of the angles should be required along with the fact that the Jacobian of the transformation is
equal to one (which is equivalent to the preservation of the volume). Since such a transformation apparently does not exist, Migdal's method fails in case of the stress-free spheroid. However, for the rigid
boundary conditions, the perturbation theory works well and provides analytical expression for the splitting of the spheroidal vibrational mode, induced by the symmetry reduction.

\acknowledgments
We would like to thank Vadim Dyadechko for useful discussions. M.O.N acknowledges the financial support by the Deutsche Forschungsgemeinschaft (project AK40/13-1, no. 506623857).
L.S. acknowledges support by the EIPHI Graduate School (contract ANR-17-EURE-0002) operated by the French National Research Agency (ANR).
The work of S.V.G. was supported in part by the U.S. Department of Energy, Office of Science, Office of Workforce Development for Teachers and Scientists (WDTS) under the Visiting Faculty Program (VFP) and, 
in part, by NSF through DMR-2100248.

\mbox{}

\section*{Data availability}
The data that support the findings of this article are not publicly available. The data are available from the authors upon reasonable request.

\end{document}